# Integration of a city GIS data with Google Map API and Google Earth API for a web based 3D Geospatial Application

**Akanbi A. K[1], Agunbiade O. Y[2]**

[1]Institute of Science & Technology, Jawaharlal Nehru Technological University,
Hyderabad 500085, A.P, India.
akanbiadeyinka@hotmail.com

[2]Department of Computer Science and Engineering, Tshwane University of Technology,
Pretoria Campus, South Africa.
agunbiadeOY@tut.ac.za

**ABSTRACT:** *Geospatial applications are becoming indispensible part of information systems, they provides detailed information's regarding the attribute data of spatial objects in real world. Due to the rapid technological developments in web based geographical information systems, the uses of web based geospatial application varies from Geotagging to Geolocation capabilities. Therefore, effective utilization of web based information system can only be realized by representing the world in its original view, where attributes data of spatial objects are integrated with spatial object and available for the user on the web, using integrated Google API and Google Earth API. In this study a city in the south-western part of Nigeria called EDE is examined and used as a case study. Using Google Map API and Google Earth API, the attribute data of the study area stored in XML databases will be integrated with the corresponding existing spatial data of the study area; to create a web based 3D geospatial application. We envisage that this system will enhance the effectiveness of web-based Geographical Information System (GIS) and the overall user experience.*

## 1. Introduction

According to ESRI (1996), Geographic Information System (GIS) is an organized collection of computer hardware, software, geographic data, and personnel designed to efficiently capture, store, update, manipulate, analyze, and display many forms of geographically referenced information [1].

The advances and development in the 21st century has brought about the warm affection for web technologies. Most application are now web based with cloud support capabilities, with Geographical Information System (GIS) not left out. Therefore, web based GIS systems has become an indispensable part of user and organization needs. Because it provides easy usage, fast sharing option, and is easily accessible to the public users.

The needs for 3D city models are growing and expanding rapidly in a different kind of fields and areas. In a steady shift from traditional 2D-GIS toward 3D-GIS, a great amount of accurate 3D city models have become necessary to be produced in a short period of time and great amount of software's become using for producing 3D city models [2].

Attribute data is very important components of any geospatial applications. With the aid of web applications these attribute data are easily accessible for the user needs across any platforms. In this research study, attribute data of the study area is integrated with Google Map and Google API for the purpose of serving a well defined attribute data for the user needs. These two applications are applied to the

With an increasing number of people living in or moving to cities, cities are getting growing and carrying more people. This shows that vital functions of people and relating urban developments must be planned, by using new techniques. In this situation, 3D city models provide effective planning, visualization tools and behave as a bridge between public and planners. But 3D city models has large amount of data. For the aim of public usage of the 3D city models on web browser, some algorithms and techniques are using nowadays [3].

Google Earth API is a commonly used code library all around the world, provided by Google, which allow users to add Google Earth Map to the web pages. So users can display 3D models inside their web pages without installing Google Earth Software by using code blocks. Similarly, Google Map API consists of code blocks that are used to modify map according to user needs [4]. API, an abbreviation of Application Program Interface, is a set of routines, protocols, that specifies how some software components should interact with each other. Google provides satellite images, road maps, terrestrial maps, and other geocoding options, because of this, in web based geographical information systems, the Google Earth and Google Map is a good source and option for users.

web page and integrated to each other using JavaScript. By adding Combo boxes that has attribute data of the buildings to the study area, the two maps are superimposed on each other. Google Earth Map is displaying 3D spatial objects and Google Map is displaying attribute data of the building using info windows. The attribute data are stored in XML databases with GeoXML function class. The ways of the study & integration steps are examined in this study.

## 2. Material Method

In this study, a web application has been prepared for Ede City located on coordinates 7°44'12.75" N   4°26'10.03" E, with the cultural heritage, which are carrying the Old Oyo Empire. Therefore, being part of the Old Oyo Empire, the city lays strategically in the South Western part of Nigeria, with Old palace (*aafin*), Town Hall, Mogaji houses (*carr*), mosques, churches etc. However, for the purpose of presentation and detailed representation of this great old empire a web page has been prepared with Google Earth & Google Map. The Old palace, Mogaji's houses, Mosques, Churches has been added to the application.

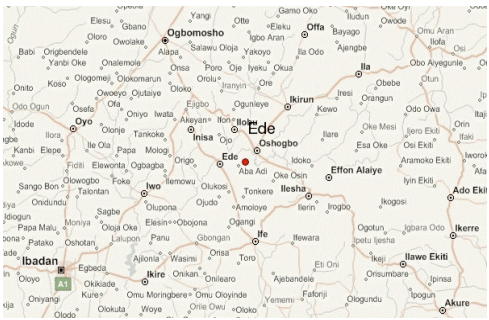

**Figure 1**: Study area Ede City in Osun, South-West Nigeria.

A typical town in the Old Oyo Empire has a palace called *aafin*, which serves as the abode for the king. The other ruling houses have a chief, who is the head of the ruling and also with his/her own mini-palace called *carr,* all with cultural heritages attached to it deeply integrated with each other.

This entire old historical site in the city is densely packed within the old empire. The figure below shows the city using a Google LandSat Image.

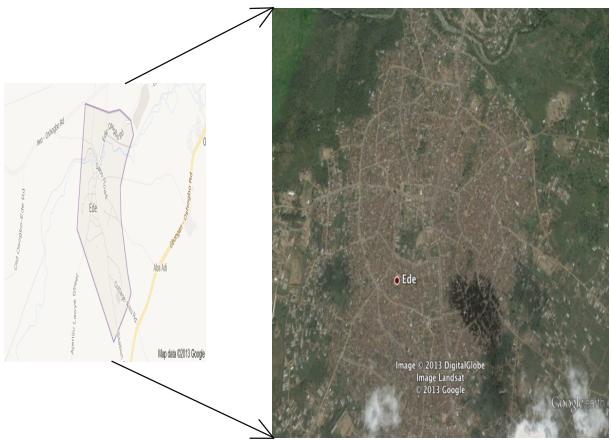

**Figure 2:** City center of Ede & LandSat Imagery of the study area.

Google Earth and Google Map API are using JavaScript language and enable users to customize applications and add maps to their own web pages [5]. For this application all the customizing and managing the data are prepared in JavaScript code blocks [6].

The JavaScript codes are embedded into the HTML codes block using the *<script>* tags.

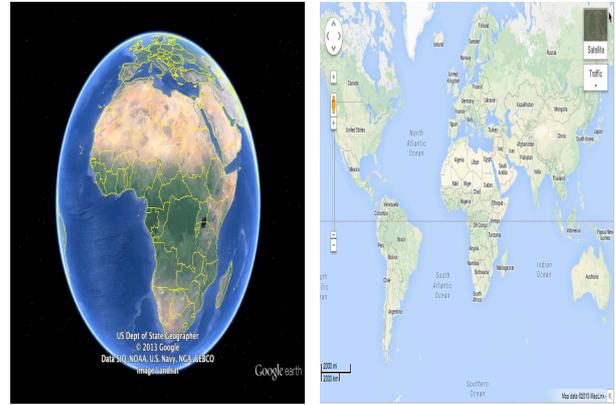

**Figure 3**: Google Earth & Google Map

Google Earth Map has been integrated into the web page for displaying the locations, directions and place marks that enhance the 3D model presentation of the study area. Also, Google Map is used to display the attribute data.

## 3. Web-Based GIS Architecture

There are basically two types of architectures for developing Web-based GIS applications: client-side, and server-side. In this research study, a server-side architecture will be utilized.

**3.1  Server-side**: This architecture allows users to submit their requests for 3D data to a Web server through the interface on the web page. The server processes the requests and returns data, which is displayed for the user on the web page. In server-side Internet GIS applications, all the complex and proprietary software, in addition to the spatial and tabular data remain on the server [7].

The advantages of this design include simplified development, deployment, and maintenance. The major benefit for server-side computing is its true cross-platform capability at the client side. Since the input and output are both generic HTML documents, no special software or plug-in are required from the user end. The user can access the desired information across different platform and different browsers. Although this model provides an interactive two-way communication between users and information providers, it is slow because each time a request is made, a new HTML document has to be generated and transferred to the client [8].

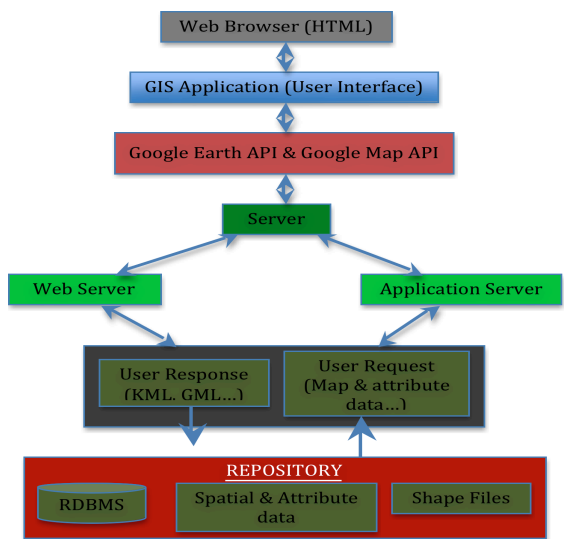

**Figure 4**: Architecture of the web based GIS

## 4. Application

The first step is defining the Google Earth Map using the coordinates of the study area where the 3D building will be shown on the map and also Google Map which will represent attributes data and detailed locational information of the 3D buildings. The code snippet for defining the map & map features are given below:

```
google.earth.createInstance
(
'earth',
function(ge) {    //Google Earth Map define
ge = ge;
map = new GMap2($('#map').get(0)); // Map define
```

**Code Snippet 1**: Google Earth & Google Map define functions

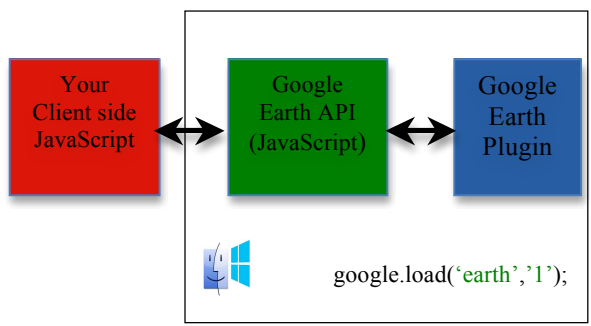

**Figure 5:** Google Earth API and JavaScript

The coordinates of the study area are defined, and can be zoomed with various map features added.

```
var la = ge.createLookAt('');
la.set(lat, lng, 10,
ge.ALTITUDE_RELATIVE_TO_GROUND, 5, 70, 300);
ge.getView().setAbstractView(la);
ge.getNavigationControl().setVisibility(true);

map.setCenter(new GLatLng(7.73687489, 4.43611944), 2);
map.addControl(new GLargeMapControl());
map.enableContinuousZoom();

map.enableDoubleClickZoom();
map.enableScrollWheelZoom();
```

**Code Snippet 2**: Defining the coordinates of the study area.

By correspondingly defining the two maps coordinates; the two maps are super-imposed and can be control at the same time. Therefore, by appending the map coordinates with click options the other map moves to the clicked location on the corresponding map. The code below shows the marker definitions.

```
window.placemark10 = DS_ge.createPlacemark('');
var point = new GLatLng(lng,lat);
var marker = new GMarker(point);
```

**Code snippet 3**: Placemark Code

The attributes showing in the marker are examined using the GeoXML and the Info Window property features. Therefore, clicking on this marker will display the attribute data of the location clicked, which was saved in the XML database between *<name>* tags.

```
var link = ge.createLink('');
var href = 'localhost/ede_3D_historicalsite/attri.xml'
link.setHref(href);
map.openInfoWindow(map.getCenter(),
document.createTextNode("name"));
```

**Code snippet 4**: Using the Info Property for the attribute data

```
<Placemark>
<name>Ede Town Hall is an
ancient hall that serves as
a point of, discussions,
functions, events and
meetings. It's at the
center of the city and
directly beside the Kings
Palace.
</name>
<name>Mosque are the places
for worship for Muslims
according to Islam
doctrine.
</name>
<styleUrl>#msnx_ylw-
pushpin</styleUrl>
<Polygon>
<tessellate>1</tessellate>
<outerBoundaryIs>
<LinearRing>
<coordinates>7.73687489253284,
4.43611944444444,
</coordinates>
</LinearRing>
</outerBoundaryIs>
</Polygon>
</Placemark>
```

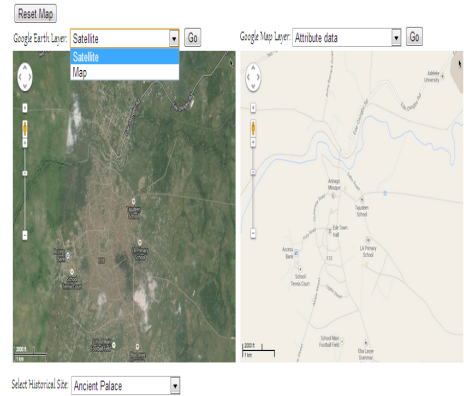

**Figure 7:** Web Application GUI

## 5. Conclusion

This web based GIS application shows how Google Map integrated with Google Earth API can be used for a comprehensive web based 3D city models. By storing the corresponding attribute data in XML database, the detailed information about a location on the 3D map is displayed which gives users a more realistic experience and information for user needs. Integrating with the Google Map allows road directions, street names and all other location information to be obtained together with the 3D buildings. This application will increases productivity and information sharing for private users, local businesses and the general public by making it easy to view, analyze, and make maps with authoritative local geographic data.

The attributes data to be displayed are added to the XML database, this enable the web page to display the attribute data for each location selected. In the figure below, the web page is shown. By selecting from the "Historical Location" drop-down box, the corresponding 3D building of the location selected is displayed, and on the right side the attribute data and location information are shown.

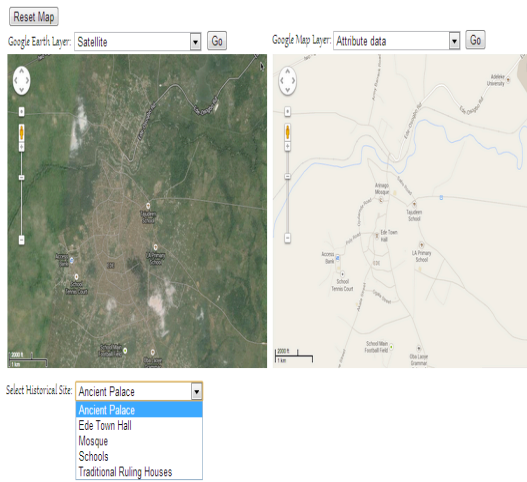

**Figure 6:** Web Application GUI

**AUTHOR PROFILE**

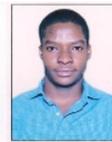

**Akanbi A. K** is a Researcher at the Institute of Science & Technology, JNTUH. He holds a Master of Science (M.Sc.) in Geospatial Science & Technology from Jawaharlal Technological University, Hyderabad, India; & Bachelor of Technology (B.Tech.) in Computer Science from Ladoke Akintola University of Technology, Ogbomoso, Nigeria. His research interests are in Ontology, Semantic Web, GIS, eCommerce, HCI, Image Processing and Artificial Intelligence. He also holds various certifications such as MCSE, MCITP, RHCSA, and RHCE from Microsoft & Redhat.

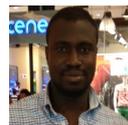

**Agunbiade O. Y** received his B.Tech in Computer Science from Ladoke Akintola University of Technology, Ogbomoso, Nigeria in 2008, and presently a master student at the Department of Computer System Engineering at Tshwane University of Technology, South Africa. His focus area lies in Image Processing, Computer Vision, Data Mining, Ubiquitous Intelligence, Artificial Intelligence, Numerical Computation and Knowledge Management.

**References**

[1]  ESRI. 1996. "Using Arc View GIS 3.2a". ESRI: USA.
[2]  Y. Takase, N. Sho, A. Sone, K. Shimiya, "Automatic Generation Of 3d City Models And Related Applications," International Archives of the Photogrammetry, Remote Sensing and Spatial Information Sciences, Vol. XXXIV-5/W10
[3]  Bo Maoa, Yifang Bana, Lars Harrie, "A multiple representation data structure for dynamic visualization of generalised 3D city models", ISPRS Journal of Photogrammetry and Remote Sensing.
[4]  Fatih SARI, Hakan KARABORK, "3D GIS Application by implementing 3D City Model with Google Earth and Google Map Integration", CIPA.
[5]  Sarle., Erdi.A., Kirtiloglu.O.S.,Kampüs Bilgi Sistemi Olusturma Çalısmaları ve Panoramik G.rüntüler,Konya Selçuk Üniversitesi Örnegi, 13.Harita Bilimsel ve Teknik Kurultayı, Ankara 2011.
[6]  Google Maps API Developer Guide. [Online]. Available: http://code.google.com/intl/tr-TR/apis/maps/index.html. [Accessed: Feb. 9, 2013].
[7]  Plewe, B, "GIS Online: Information Retrieval, Mapping, and the Internet". On Word Press: Santa Fe, NM, 1997.
[8]  Ran, B. and B.P. Chang. 1999. "Architecture Development for Web-Based GIS, Application in Transportation". Transportation Research Board Annual Meeting. Revised November 17, 1998.